\documentclass[12pt]{iopart}

\usepackage{subcaption}
\usepackage{graphicx}
\usepackage{wrapfig}
\usepackage{cite}
\usepackage{diagbox}
\usepackage{changepage}
\begin{document}

\title[A New Torsion Balance for the Search of Long-range Interactions]{A New Torsion Balance for the Search of Long-range Interactions Coupling to Baryon and Lepton Numbers}

\author{Ramanath Cowsik, Dawson Huth, Tsitsi Madziwa-Nussinov}

\address{McDonnell Center for the Space Sciences\\
and Physics Department,\\
Washington University in St. Louis\\
One Brookings Dr. \\
St. Louis, MO 63130}
\ead{cowsik@physics.wustl.edu}
\vspace{10pt}
\begin{indented}
\item[]March 2021
\end{indented}

\begin{abstract}
We have developed a torsion balance with a sensitivity about ten times better than those of previously operating balances for the study of long range forces coupling to baryon and lepton numbers. We present here the details of the design and expected characteristics of this balance. Operation of this balance for a year will also result in improved bounds on long range interactions of dark matter violating Einstein’s equivalence principle.
\end{abstract}

%
%
%
%
%

\section{Introduction}
For over a century experimental efforts have probed Einstein's theory of General Relativity (GR) finding agreement with all its theoretical predictions \cite{Will_2014, PDG_2020}. Even though its original formulation was based only on Einstein's Equivalence Principle (EEP) and general covariance, the Strong Equivalence Principle (SEP), which requires gravitational self-interactions of massive bodies like stars and planets also obey the tenets of EEP, was an unstated requirement. Lunar laser ranging and observations of pulsars in multiple stellar systems have tested SEP as well \cite{Williams_2004, Archibald_2018}. Recent analysis of a SPARC sample of galaxies by Chae et al. \cite{Chae_2020} however has cast doubts on the validity of the SEP. If confirmed this is the first indication that we need to modify GR. Similarly, the Standard Model of Particle Physics (SM) is a non-abelian gauge theory with the symmetry group U(1)$\times$SU(2)$\times$SU(3), which has also been eminently successful in accommodating an extensive set of experimental findings probing its validity. However, the model, while explaining elegantly the phenomena related to strong, electromagnetic, and weak interactions, is not able to provide as yet an acceptable candidate particle that can serve as dark matter. Nor is it able to include a quantized version of gravitation within its framework. There have been several attempts at extension of SM, which predict the existence of new forces \cite{Damour_2012, Fayet_1990, Fischbach_1986} that will appear as a signal in apparent violation of EEP in our experiment . Our torsion balance is designed specifically to search for such forces and characterize them.

Some of the best modern tests of EEP constrain the E\"{o}tv\"{o}s parameter $\eta$ which describes the validity of the universality of free fall (UFF), also called the Weak Equivalence Principle (WEP). The E\"{o}tv\"{o}s parameter encodes the difference in the accelerations $a$ between two objects composed of materials $A$ and $B$ subject to the same gravitational field and is defined as
\begin{equation}
    \eta(A,B) \equiv \frac{2(a_A - a_B)}{a_A + a_B}.
\end{equation}
Terrestrial experiments using rotating torsion balances have placed upper bounds on composition-dependent forces in terms of the E\"{o}tv\"{o}s parameter $\eta(\mathrm{Be,Ti}) = (0.3\ \pm\ 1.8) \times 10^{-13}$ \cite{Schlamminger_2008} and $\eta(\mathrm{Be,Al}) = (-0.7\ \pm\ 1.3) \times 10^{-13}$ \cite{Wagner_2012}. More recently torsion balance tests of the WEP have used chiral test masses probing violations of gravitational parity \cite{Bargueno_2015, Dass_1976} reporting $\eta(\mathrm{left,right}) = [-1.2\ \pm\ 2.8(\mathrm{stat})\ \pm\ 3.0(\mathrm{syst})] \times 10^{-13}$ \cite{Zhu_2018}. The first results from the MICROSCOPE space based mission have reported $\eta(\mathrm{Ti,Pt}) = [-1\ \pm\ 9(\mathrm{stat})\ \pm\ 9(\mathrm{syst})] \times 10^{-15}$ \cite{MICROSCOPE} and limits on new long-range forces have been placed with these results \cite{Fayet_2018, Fayet_2019}. Astronomical observations of the pulsar J0337+1715 in a triple star system with two white dwarfs was also recently used to constrain $\eta$ in the regime of the Strong Equivalence Principle (SEP), where contributions from massive self-gravitating test bodies can no longer be neglected. A. M. Archibald et al. \cite{Archibald_2018} analyzed timing observations of the pulses from the pulsar over a six-year period showing that the relative accelerations of the white dwarfs and the neutron star varied by no more than a fraction $\sim2.6\times10^{-6}$ of their mean accelerations or $\eta_{SEP} \sim 1.7 \times 10^{-5}$.

The possible existence of a 'dark/hidden sector' of particles which are neutral to the forces of the SM has been a leading motivation for Beyond Standard Model (BSM) physics; EEP experiments are capable of probing the parameter spaces of these models \cite{Carroll_2009}. Despite the sensitive bounds on $\eta$ shown above plenty of untouched parameter space exists for future EEP violation searches to explore in the context of BSM physics and constrain properties of new particles of interest such as dark photons and Weakly Interacting Massive Particles (WIMPs) \cite{Fayet_2021, Fabbrichesi_2021, Arcadi_2018}. 

The work presented here covers a recent 'pilot experiment' of the operation of a long-period torsion balance instrument sensitive to long-range forces coupling to baryon and lepton numbers. We discuss the subsequent instrument upgrades and the design of a new torsion balance developed to enhance our instrument's sensitivity to these forces. We conclude with the expected response of the new balance to forces violating the WEP and the prospects for placing lower bounds on $\eta$.

\section{Pilot Experiment}
The design of the pilot long-period torsion balance shown in Fig. 1 follows the classic design concepts developed by Dicke \cite{Dicke_1964}, Braginsky \cite{Braginsky_1972}, and the more recent works of the E\"{o}t-Wash group \cite{Wagner_2012} and Zhu et al.  \cite{Zhu_2018}. The balance bob has four-fold azimuthal symmetry with 14.33 g test masses composed of Al and SiO$_{2}$. This symmetry significantly reduces the bob's coupling to gravitational gradients. The composition of the test masses were chosen to give large differences in baryon number per amu $(B/\mu)$, lepton number per amu $(L/\mu)$, and $(B-L)/\mu$ thereby enhancing the bob's sensitivity to equivalence principle (EP) violating forces which couple to these charges \cite{Damour_1996}. Values for these parameters with respect to this pilot experiment, the E\"{o}t-Wash group balance, and the new balance we have constructed can be found in Table 1. The composition dipole generated by these charge differences is subject to the gravitational field of the Sun and that of the dark matter halo centered about our Galactic Center. It is expected that any WEP violating force associated with those gravitational fields will exert a torque on the balance bob with a period of the solar or sidereal day, respectively. Data acquisition with this balance lasted approximately six months and yielded $\sim115$ continuous days of data used in analysis. The balance bob's long natural period combined with the low frequency of the diurnal or sidereal signal cause long strings of uninterrupted data acquisition to be critical for the success of this instrument. This pilot experiment shows that collecting data of this kind is indeed possible. 

\begin{figure}
  \begin{center}
    \includegraphics[width=0.42\textwidth]{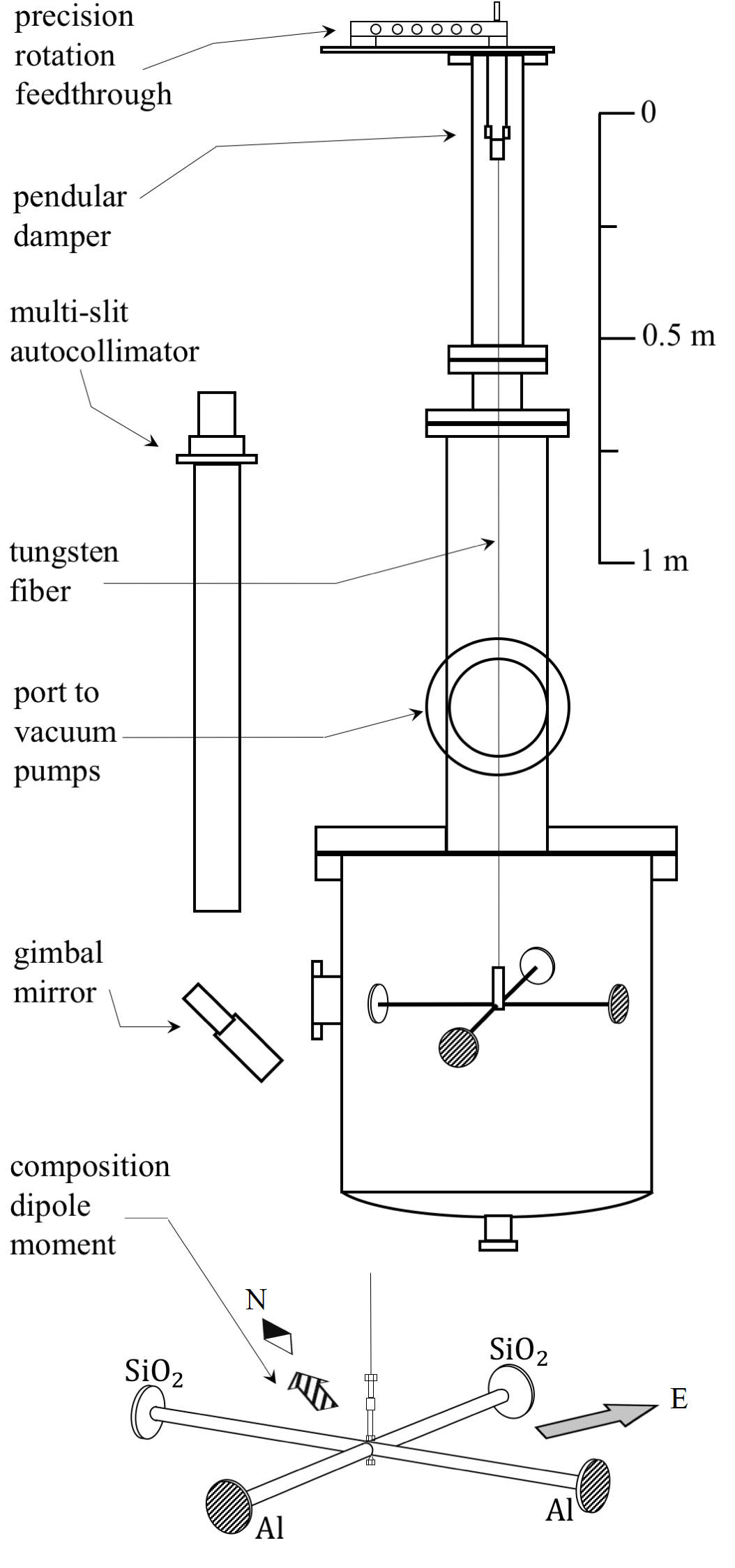}
  \end{center}
  \caption{Line Drawing of the instrument used in the pilot experiment}
\end{figure}

In principle, a torsion balance of this design is significantly more sensitive to WEP violating forces than the balance of the E\"{o}t-Wash group but some noise sources were insufficiently addressed. Our instrument's pendular damper did not operate as efficiently as expected and large noise contributions in the low frequency regime of the balance's operation limited our final result. Thorough studies of noise were performed to understand the instrument's response to various environmental parameters. Environmental isolation techniques have been used to address these noise sources like installing magnetic shielding and a thermal gradient control system.

\section{Noise Diagnostics and Remediation}
We have studied the response of the instrument to various noise sources such as fluctuations in the Earth's magnetic field, temperature gradients, and atmospheric pressure variations. In each noise study a length of $\sim 2$ weeks of the pilot balance's position and environmental data were collected. These data were then used in cross-correlation analysis where normalized correlation coefficients are calculated between the balance position and various environmental parameters under study. These coefficients are given by
\begin{equation}
    R_{xy}(\tau) = \frac{1}{T}\int_{-T}^{T} \frac{(x(t + \tau) - \mu_x) (y(t) - \mu_y)}{\sigma_x \sigma_y}dt,
\end{equation}
where $R_{xy}$ is the normalized correlation coefficient between two sets of data $x$ and $y$ of length $T$ calculated for some time lag $\tau$. These data sets have means $\mu$ and standard deviations $\sigma$. This analysis showed that correlations with the Earth's magnetic field and variations in temperature gradients are the most significant noise contributions with normalized correlation coefficients of 0.4 and 0.3 respectively. Pressure variations yielded coefficients at the level of 0.01 and the coefficients for absolute temperature were also very low. Plots of the correlation coefficients as functions of time lag for the largest contributors are shown in Fig. 3.

\begin{figure}
  \begin{center}
    \begin{subfigure}{.8\textwidth}
    \centering
    \includegraphics[width=\textwidth]{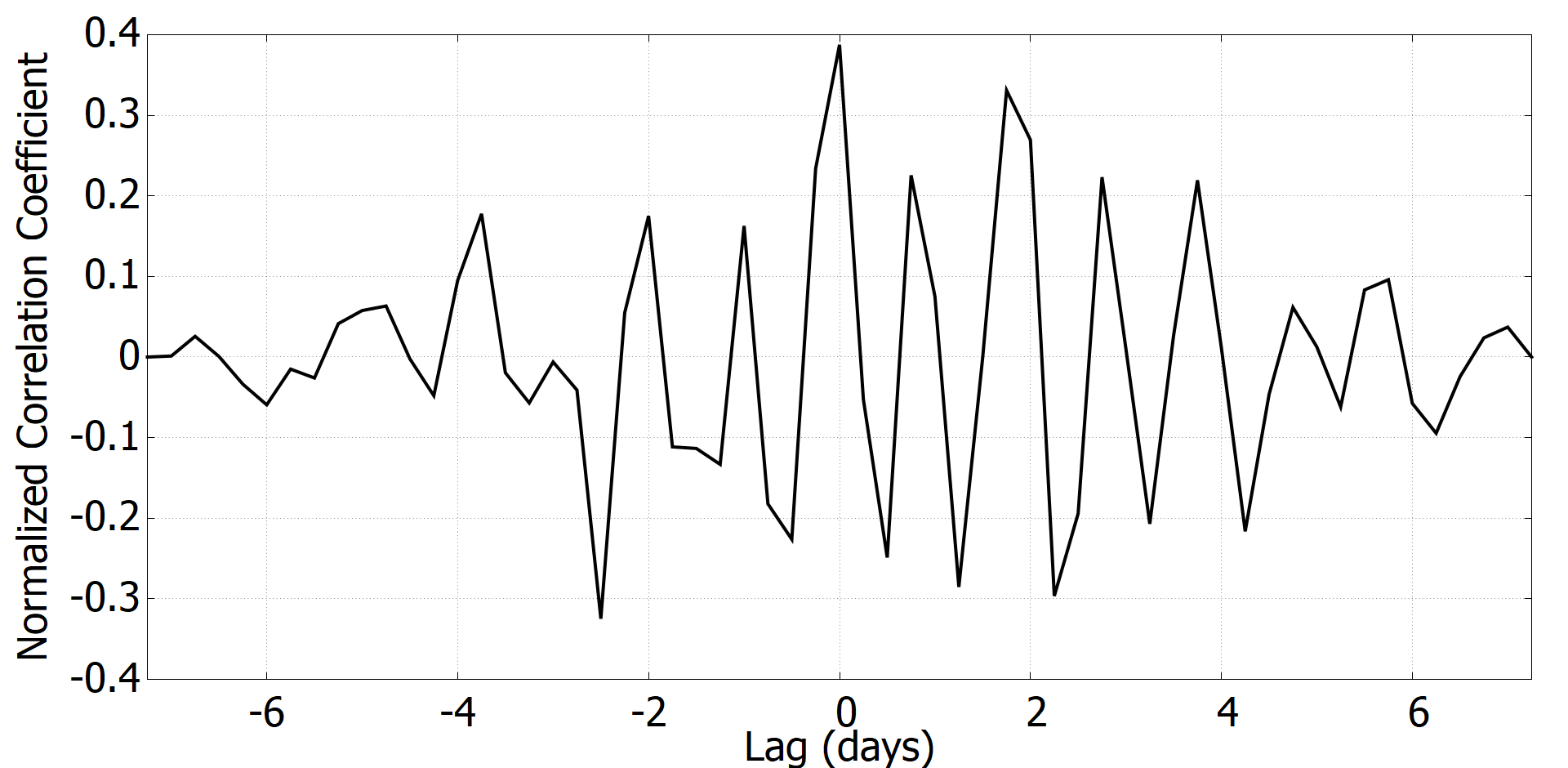}
    \end{subfigure}
    \begin{subfigure}{.8\textwidth}
    \centering
    \includegraphics[width=\textwidth]{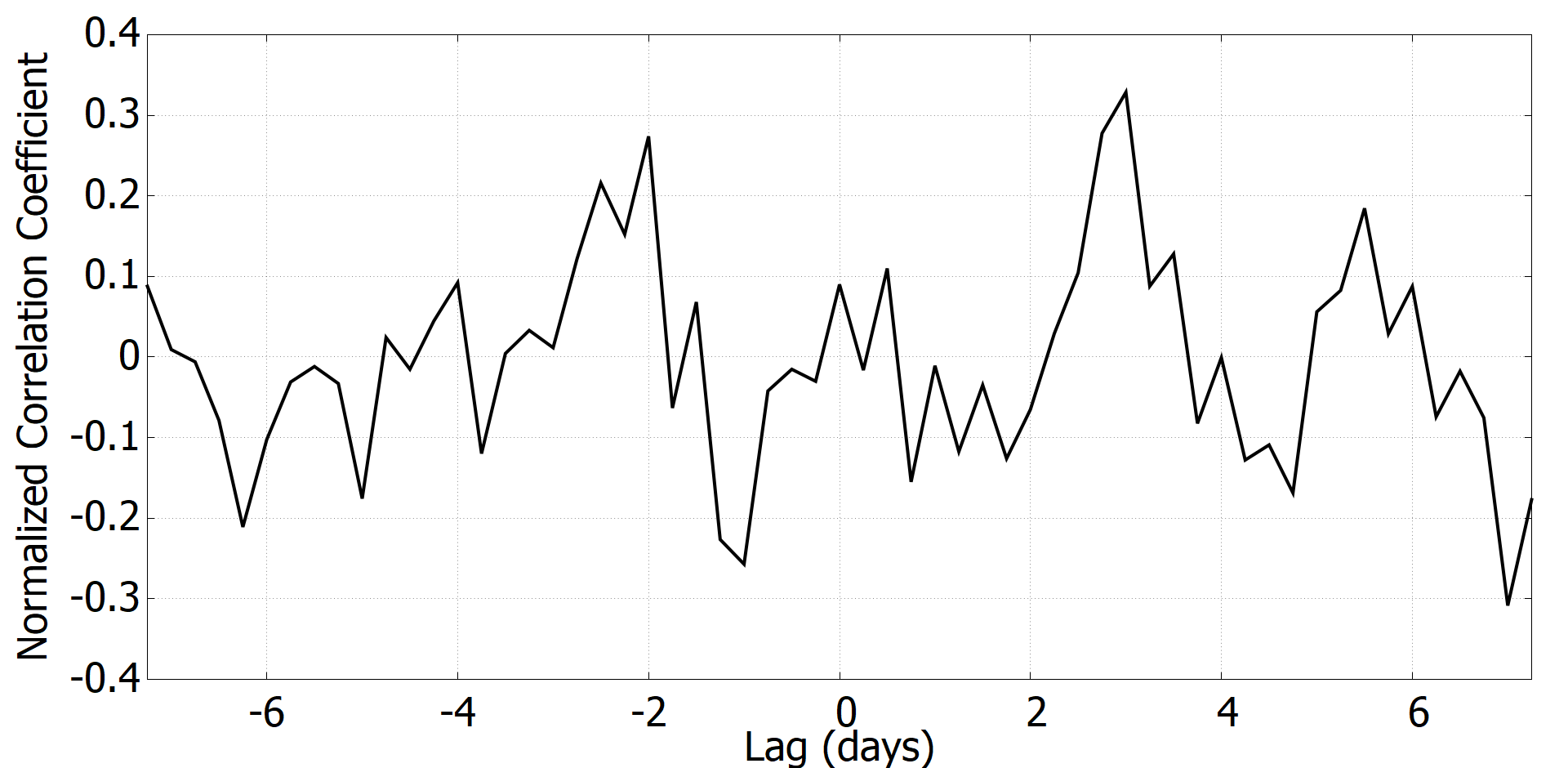}
    \end{subfigure}
  \end{center}
  \caption{Plots of correlation coefficients between balance position and the Earth's magnetic field (top) and temperature gradients (bottom) in lag increments of 6 hours. }
\end{figure}

Based on these studies we added magnetic shielding around the chamber and developed a water circulation system for evening out thermal gradients. For magnetic shielding two layers of a woven low-carbon steel called G-IRON are used to completely surround the part of the vacuum chamber containing the balance. The woven structure of the material allows for sufficient pliability to be wrapped around the chamber without significant losses to magnetic permeability. Testing of this material showed good shielding performance with two concentric layers reducing the magnitude of the Earth's field measured with a magnetometer by about a factor of 40. To suppress the thermal gradients a novel water circulation system was developed. Six air-to-water heat exchangers will surround the balance and be connected in series. A low power, ultra-quiet water pump connected to a large water tank will slowly circulate water through these heat exchangers evening out thermal gradients and reducing the amplitude of the diurnal thermal wave which propagates through the instrument. When the chamber is fully insulated with two layers of polyethylene foam, with the heat exchangers between the two layers, convection is the dominant heat transfer mechanism. With these modifications we expect significant reduction of thermal noise in the system.

\section{Design of the New Balance}
We are interested in detecting long-range forces which couple to baryon and lepton numbers so choosing test-body materials which maximize the contrast in these quantities is critical to a sensitive measurement of $\eta$. To this end a new balance was designed with Cu and ultra-high molecular weight polyethylene (UHMWP) as test mass materials. We ensured UHMWP is vacuum compatible by placing a sample with high surface area in a vacuum chamber separate from our instrument. The sample was pumped down to $\sim 10^{-8}$ torr within a day and reached a few parts in $10^{-9}$ torr after a few days giving a sufficient level of vacuum for our experiment without a prolonged outgassing period. Compared to the materials of the pilot balance this choice increases $\Delta (B/\mu)$ by over an order of magnitude; similarly $\Delta (L/\mu)$ and $\Delta (B-L/\mu)$ are enhanced by a factor of $\sim 6-7$ and these values can be compared with the E\"{o}t-Wash balances in Table 1. Perusal of Table 1 offers a comparison of the mechanical properties of the prototype, proposed, and E\"{o}t-Wash instruments.

The balance is designed with a ring-shaped geometry where each semicircle has the same mass and the same first and second order moments. The copper semicircle is comprised of four 90$^{\circ}$ arcs with two arcs stacked vertically on each quadrant of the semicircle separated by $\sim 3.8$ cm. The UHMWP semicircle has two 90$^{\circ}$ arcs joined at the ends. The UHMWP semicircle is also covered with Cu foil to prevent the accumulation of patch charges and the entire assembly is joined together with conductive epoxy to make a circular ring. The higher azimuthal symmetry reduces couplings to gravitational gradients, which induce spurious torques on the balance, and also removes any preferential direction it may be deflected by radiometric flow in contrast with four-fold symmetric designs. The total mass of the balance is 530 g, which provides larger coupling of any WEP violating forces to the balance while reducing the response to torques suffered by the balance from radiometric flow inside the vacuum chamber. This flow is induced by temperature gradients across the vacuum chamber which we expect had a significant effect on the pilot balance. A photograph of the assembled balance is shown in Fig. 3.

\begin{table}[ht!]
\footnotesize
\begin{adjustwidth}{-0.5in}{-0.5in}
\begin{center}
\bgroup
\def\arraystretch{1.10}
\begin{tabular}{|c|c|c|c|}
\hline
Characteristics & E\"{o}t-Wash & Pilot & Current \\
\hline
\hline
Materials & \diagbox[]{Be-Ti \cite{Schlamminger_2008}}{Be-Al \cite{Wagner_2012}}  & Al-SiO$_2$ & Cu-C$_2$H$_4$\\
\hline
Total Mass (g) & 70 & 72 & 530 \\
\hline
Tine Length/Radius (cm) & 2.01 & 25 & 25\\
\hline
Moment of Inertia (g cm$^2$) & $3.78\times10^2$ & $3.75\times10^4$ & $2.96\times10^5$\\
\hline
Fiber Length (m) & 1.07 & 1.67 & 1.67\\
\hline
Fiber Section & 20 $\mu$m dia. & 18 $\mu$m dia. & 15 $\times$ 150 $\mu$m$^2$\\
\hline
Torsion Constant (dyne cm rad$^{-1}$) & $2.34 \times 10^{-2}$ & $9.03 \times 10^{-3}$ & $7.93\times 10^{-2}$\\
\hline
Natural Period (s) & 798 & 12800 & 12100\\
\hline
Signal Torque (dyne cm) & $3.98 \times 10^{-12}$ & $5.07 \times 10^{-11}$ & $3.98 \times 10^{-10}$\\
\hline
Expected Deflection (rad) & $1.66 \times 10^{-10}$ & $5.62 \times 10^{-9}$ & $5.02 \times 10^{-9}$\\
\hline 
Nyquist Torque (dyne cm) & $3.14 \times 10^{-12}$ & $6.34 \times 10^{-12}$ & $1.84 \times 10^{-11}$\\
\hline
Signal-to-Noise Ratio ($\tau_S / \tau_{Ny}$) & $1.24$ & $7.99$ & $21.63$\\
\hline
\hline
Balance Composition Characteristics & & &\\
\hline
$\Delta(B/\mu)$ & \diagbox[]{$2.2429\times10^{-3}$}{$2.0361\times10^{-3}$} & $1.2932\times10^{-4}$ & $2.2351\times10^{-3}$\\
\hline
$\Delta(L/\mu)$ & \diagbox[]{$1.5763\times10^{-2}$}{$3.7967\times10^{-2}$} & $1.6667 \times 10^{-2}$ & $1.1398 \times 10^{-1}$\\ 
\hline
$\Delta(B-L/\mu)$ & \diagbox[]{$1.3337\times10^{-2}$}{$3.5928\times10^{-2}$} & $1.8717 \times 10^{-2}$ & $1.1621 \times 10^{-1}$\\
\hline
\hline
Charge Composition of Sun & $(B/\mu)$ & $(L/\mu)$ & $(B-L/\mu)$\\
\hline
& 0.9943 & 0.8493 & 0.1450\\
\hline
\end{tabular}
\egroup
\end{center}
\end{adjustwidth}
\caption{Comparison of characteristics of our prototype, proposed, and the E\"{o}t-Wash instruments. Values for expected deflection, signal torque, and Nyquist torque assume a WEP violation at the level of $\eta \sim 10^{-13}$ and an observation time of $10^7$ s. The charge characteristics of the Sun assuming a composition of $71\%$ H and $29\%$ He are shown for reference.}
\end{table}

\begin{figure}
  \begin{center}
    \begin{subfigure}{.4\textwidth}
    \centering
    \includegraphics[width=\textwidth]{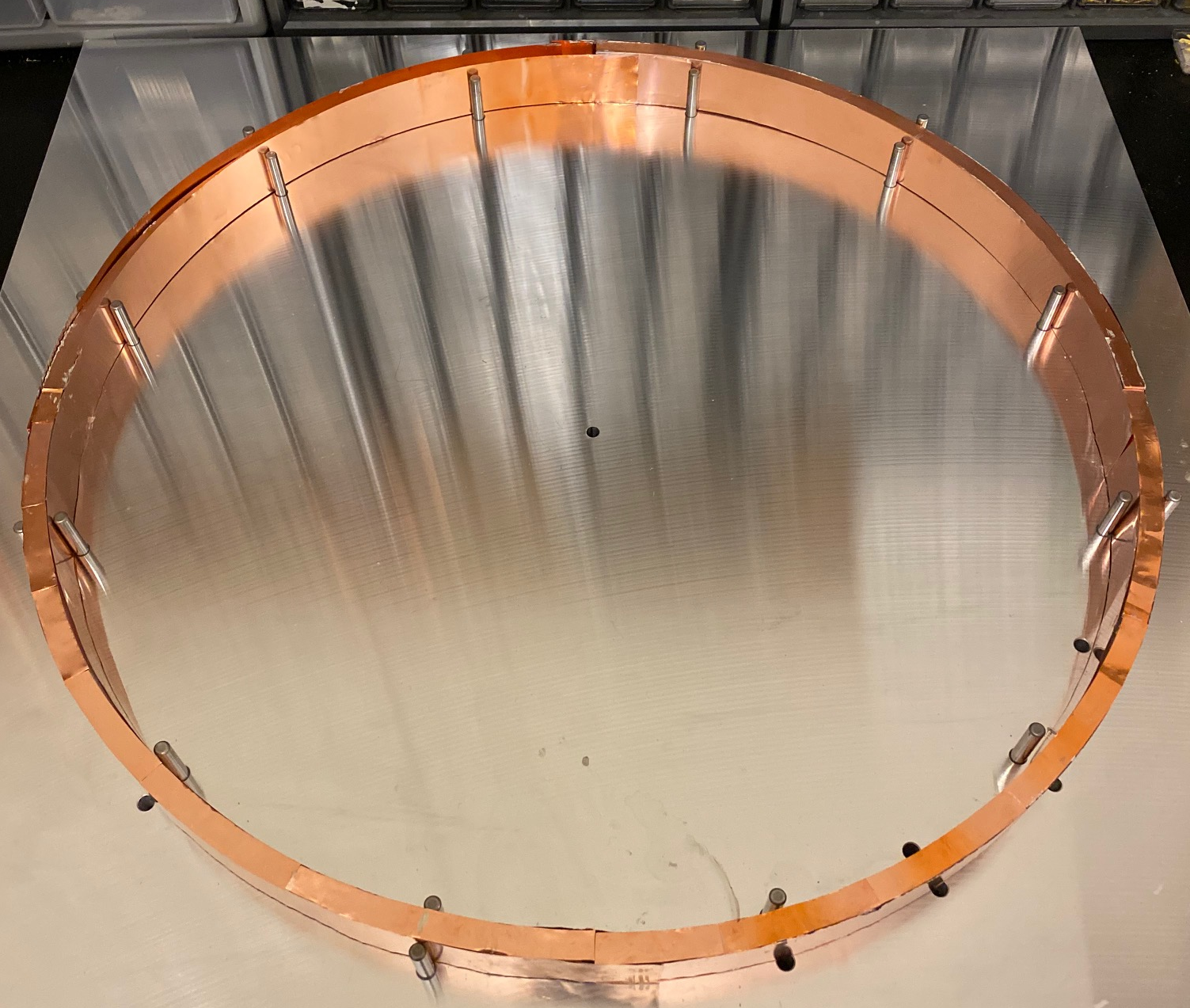}
    \end{subfigure}
    \begin{subfigure}{.4\textwidth}
    \centering
    \includegraphics[width=\textwidth]{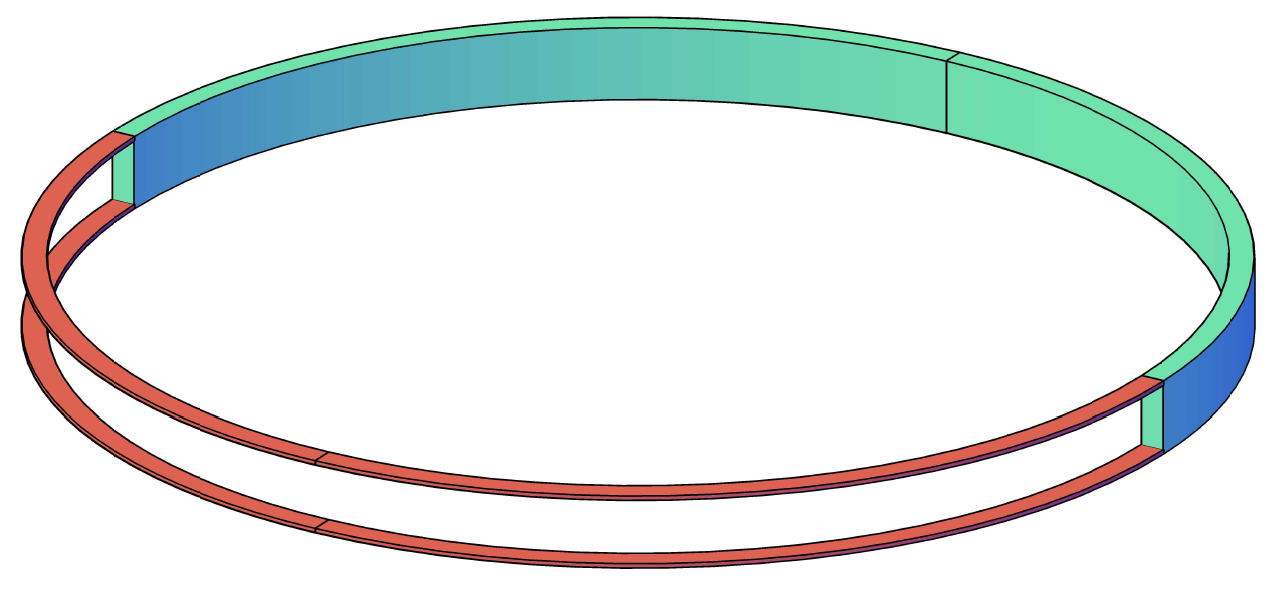}
    \end{subfigure}
  \end{center}
\caption{The constructed Cu-UHMWP balance on an aluminum jig used during assembly (left) and a 3D rendering of the balance test masses (right).}
\end{figure}

    A thin tungsten fiber is used to suspend the balance because of tungsten's high tensile strength and large Q factor. The power spectral density of data from the pilot experiment is shown in Fig. 4 with a Lorentzian fit parameterized by a Q of 800 which is limited by the length of the data set. Tungsten has been shown to offer a Q factor up to 6000 by the E\"{o}t-Wash group \cite{Schlamminger_2008}. A data set much longer than the 100 days of observations that we had is needed to establish such a high value of Q for our balance with a natural period of $\sim 12,500$ seconds. A fiber of rectangular cross-section is used rather than a circular section. Circular section fibers with radius $r$ have a torsion constant proportional to $r^4$ while a rectangular section is proportional to $a\cdot b^3$ where $a$ is the width and $b$ is the thickness of the tungsten strip. The rectangular geometry allows for a sufficiently large cross-sectional area needed to support the more massive balance while only increasing the torsion constant of the fiber by a factor of $\sim 8.8$ rather than 60 for a circular section. The tungsten strip is attached to the disc of the pendular damper, which is suspended by a tungsten fiber of circular cross section, and as such reduces the sensitivity of the torsion balance to tilts. For $\eta = 10^{-13}$ this leads to an expected deflection of $5 \times 10^{-9}$ rad. The new balance has a significantly larger signal-to-noise ratio (SNR) as compared with earlier balances (See Table 1).

\begin{figure}
    \begin{center}
    \includegraphics[width=0.9\textwidth]{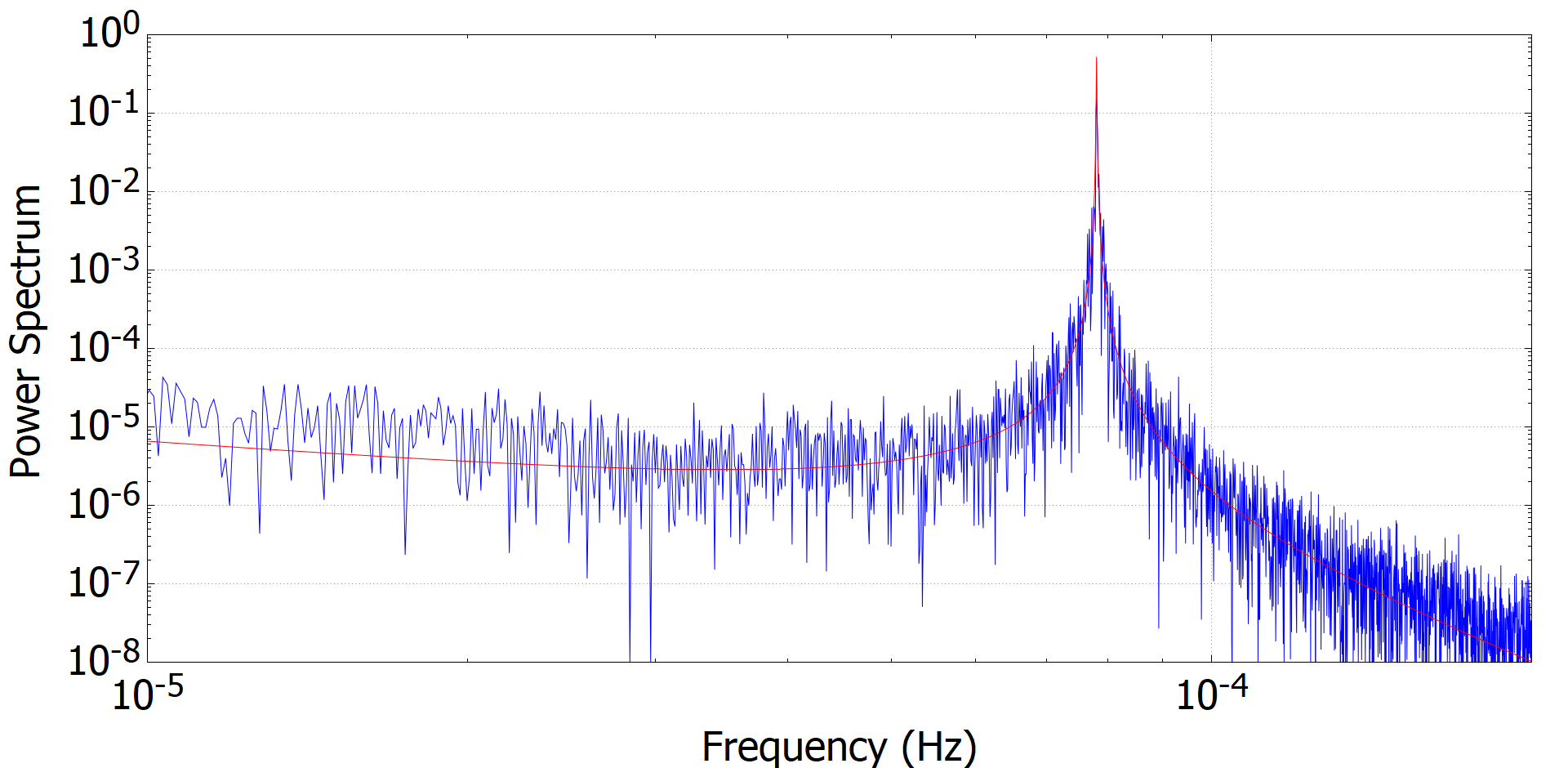}
    \end{center}
    \caption{The power spectral density of 115 days of balance position data with a Lorentz fit parameterized with Q = 800. The peak is at the natural frequency of the pilot balance $\sim 7.8 \times 10^{-5}$ Hz.}
\end{figure}

\section{Closing Remarks}
Terrestrial experiments still hold much promise for measuring violations of the EEP to greater precision and the various motivations for new physics beyond the Standard Model fuel these efforts. In this work we have described past and current works searching for EP violations including a pilot experiment of a long-period torsion balance instrument. The lessons learned from this prototype have given us insight into where improvements need to be made for better environmental isolation and reduce the noise in the torsion balance, which is projected to be in operation by September-December 2021. Many of these changes have been implemented or are actively being developed and a measurement of $\eta$ at the level of $10^{-13}$ or lower can be made. 

\section{Acknowledgements}
We would like to thank the NSF for initial funding of this project, followed by funding from the McDonnell Center for the Space Sciences. We also recognize the earlier contributions of Michael Abercrombie, Adam Archibald, Maneesh Jeyakumar, Nadathur Krishnan, and Kasey Wagoner towards this effort.

\section*{References}
\bibliographystyle{unsrt}
\bibliography{EQP2020noteBib}

\end{document}